\documentclass[3p,times,twocolumn]{elsarticle}
 
\usepackage{graphicx}
\usepackage{amssymb}

\usepackage[figuresright]{rotating}

\def\nin{\noindent}
\def\be{\begin{equation}}
\def\ee{\end{equation}}
\def\bea{\begin{eqnarray}}
\def\eea{\end{eqnarray}}

\def\la{\langle}
\def\ra{\rangle}
\def\ms{\overline{\rm MS}}

\def\t{\tilde }
\def\Lam{\Lambda}
\def\ga{\gamma}
\def\slashed{\not\!\!}
\def\nn{\nonumber}

\def\qq{\langle \bar q q \rangle} 

\begin{document}

\title{ Quark Condensate from Renormalization Group Optimized Spectral Density}
 \cortext[cor0]{Talk given at 18th International Conference in Quantum Chromodynamics (QCD 15,  30th anniversary),  
 29 june - 3 july 2015, Montpellier - FR}
 \author[label1]{J.-L.~Kneur\fnref{fn1}}
 \author[label1]{A.~Neveu}
\address[label1]{Laboratoire Charles Coulomb (L2C), UMR 5221 CNRS-Universit\'e de Montpellier, Montpellier, France}
   \fntext[fn1]{Speaker}

\pagestyle{myheadings}
\markright{ }
\begin{abstract}
Our  renormalization group consistent variant of optimized perturbation, RGOPT, is used
 to calculate the nonperturbative QCD spectral density of the Dirac operator and the related
chiral quark condensate $\la \bar q q \ra$, for $n_f=2$ and $n_f=3$ massless
quarks. Sequences of approximations at two-, three-, and 
four-loop orders are very stable and give 
$\la\bar q q\ra^{1/3}_{n_f=2}(2\, {\rm GeV}) = -(0.833-0.845) \bar\Lam_2 $, and 
$ \la\bar q q\ra^{1/3}_{n_f=3}(2\, {\rm GeV}) =  -(0.814-0.838) \bar\Lam_3 $ where the range
is our estimated theoretical error and $\bar\Lam_{n_f}$ the basic QCD scale in the $\ms$-scheme.  
We compare those results with other recent determinations (from lattice calculations and spectral sum rules).
\end{abstract}
\maketitle
\section{Introduction}
\nin
The chiral quark condensate $\la \bar q q \ra$ is
a main order parameter of spontaneous chiral symmetry breaking, $SU(n_f)_L\times SU(n_f)_R \to SU(n_f)_V$ 
for $n_f$ massless quarks. As such it is intrinsically nonperturbative, indeed 
vanishing at any finite order of (ordinary) perturbative QCD in the chiral limit. 
For $m_q\ne 0$ the famous GMOR relation~\cite{GMOR}, e.g. for two flavours:
\be
F^2_\pi\,m^2_\pi = -(m_u +m_d) \la \bar u u \ra +{\cal O}(m^2_q),
\label{GMOR}
\ee
relates the condensate with the pion mass and decay constant $F_\pi$.
At present the light quark masses $m_{u,d,s}$ determined from lattice simulations (see \cite{LattFLAG} for a review) or
using spectral sum rules~\cite{SVZSSR} (see e.g.~\cite{qqSR},\cite{qqSRlast})  
can give from (\ref{GMOR}) an indirect precise determination of the condensate.
But more direct ``first principle'' determinations are highly desirable. Analytical determinations were attempted in various models 
and approximations, starting with the Nambu and Jona-Lasinio model~\cite{NJL,NJLrev}, 
or more recently Schwinger-Dyson equations~\cite{D-S,qqbar_link} typically.
Lattice calculations have also determined 
the quark condensate by different approaches, in particular by computing the spectral density
of the Dirac operator~\cite{qqlattSD,SDlatt_recent}, directly related to the quark condensate via the
Banks-Casher relation~\cite{BanksCasher,SDgen,SDchpt2}. However, while many lattice results 
are statistically precise, they rely on extrapolations to the chiral limit, 
often using chiral perturbation theory~\cite{chpt} for this purpose. 
On phenomenological grounds, a significant suppression of the three-flavor case with respect to the two-flavor case 
has been claimed~\cite{qqflav}, which may be attributed to the relatively large explicit chiral symmetry breaking from the
strange quark mass. Moreover the convergence of chiral perturbation for $n_f=3$ appears less good, with 
different lattice results showing rather important discrepancies~\cite{LattFLAG}).\\ 
Our recently developed renormalization group optimized perturbation (RGOPT) 
method~\cite{rgopt1,rgopt_Lam,rgopt_alphas}  provides analytic sequences of  
nonperturbative approximations with a non-trivial chiral limit. We report here on the 
RGOPT calculation of the quark condensate using the spectral density, performed in \cite{rgopt_qq}. 
\section{Spectral density and the quark condensate}
We consider the (Euclidean) Dirac operator of eigenvalues $\lambda_n$
and eigenvectors $u_n$~\cite{BanksCasher,SDgen},
\be
 {\rm i}\,{\slashed D}\: u_n(x) =  \lambda_n\: u_n(x);\;\;\;{\slashed D} \equiv {\slashed{\partial}} + g\, {\slashed{A}},
\ee
where ${\slashed D}$ is the covariant derivative operator and $A$ the gluon field. Except for zero modes,
the eigenvectors
come in pairs $\{ u_n(x); \gamma_5\,u_n(x) \}$, 
with ($A$-dependent) eigenvalues $\{\lambda_n; -\lambda_n \}$.
On a lattice with finite volume $V$ the spectral density is by definition
\be
\rho(\lambda) \equiv \frac{1}{V} \la \sum_n \delta(\lambda-\lambda_n^{[A]})\ra\, ,
\label{rho_discrete}
\ee
where $\delta(x)$ is the Dirac distribution and 
$\la \cdots \ra$ designates averaging over the gauge field configurations, 
$\la \ra = \int [{\rm d}A] \prod_{i=1}^N \det ({\rm i}{\slashed D}+m)$.
The quark condensate is
\be
\frac{1}{V} \int_V {\rm d}^4 x \la \bar q(x) q(x)\ra = -2\, \frac{m}{V} \sum_{\lambda_n >0} 
\frac{1}{\lambda^2_n+m^2}.
\ee
Now when $V\to\infty$ the operator spectrum becomes dense, so with $\rho(\lambda)$ defining the spectral density,
\be
\la \bar q q  \ra = -2\, m \int_0^\infty {\rm d}\lambda \frac{\rho(\lambda)}{\lambda^2+m^2}\, .
\label{SD}
\ee
The Banks-Casher relation~\cite{BanksCasher} is the $m\to 0$ limit, giving
the condensate in the chiral limit as
\be
\lim_{m\to 0} \la \bar q q  \ra = -\pi \rho(0).
\label{BC}
\ee 
Note from the defining relations 
(\ref{rho_discrete}), (\ref{SD}) that
\be
\rho(\lambda)=\frac{1}{2\pi} \left[\qq({\rm i}\lambda-\epsilon)-
\qq({\rm i}\lambda+\epsilon)\right]|_{\epsilon\to 0}\, ,
\label{disc}
\ee
{\it i.e.} $\rho(\lambda)$ is determined by the discontinuities of $\qq(m)$ across the imaginary axis.
When $m\ne 0$, 
$\la \bar q q(m)  \ra$ has a standard QCD perturbative series
expansion, known to three-loop order at present, and its discontinuities are simply given
by perturbative logarithmic ones.  The $\lambda\to 0$ limit, relevant
for the true chiral condensate, trivially lead to a 
vanishing result~\cite{qqbar_link}. But as we recall below a crucial feature of RGOPT is to circumvent this,  
giving a nontrivial result for $\lambda \to 0$.
\section{RG optimized perturbation (RGOPT)}
The OPT key feature is to 
reorganize the standard QCD Lagrangian by "adding and subtracting" an {\em arbitrary} (quark) mass term, treating  
one mass piece as an interaction term.  To organize this
systematically at arbitrary perturbative orders, it is convenient to introduce 
a new expansion parameter $0<\delta<1$, interpolating between ${\cal L}_{free}$ and 
${\cal L}_{int}$, so that the mass 
 $m_q$ is traded for an arbitrary trial parameter.
This is perturbatively equivalent
to taking any standard perturbative expansions in $g\equiv 4\pi\alpha_S$, after renormalization, reexpanded  
in powers of $\delta$, so-called $\delta$-expansion~\cite{delta} after substituting:
\be m_q \to  m\:(1- \delta)^a,\;\; g \to  \delta \:g\;.
\label{subst1}
\ee
Note in (\ref{subst1}) the exponent $a$ reflecting a possibly more general interpolation, 
but as we recall below $a$ is uniquely fixed from requiring~\cite{rgopt_Lam,rgopt_alphas} 
consistent renormalization group (RG) invariance properties.  
Applying (\ref{subst1}) to a given perturbative expansion for a physical quantity $P(m,g)$, reexpanded in $\delta$ 
at order $k$, and taking {\em afterwards} the $\delta\to 1$ limit to recover the original {\em massless} theory,  
leaves a remnant $m$-dependence at any finite $\delta^k$-order. 
The arbitrary mass parameter $m$ is most conveniently fixed by an optimization (OPT) prescription:
\be
\frac{\partial}{\partial\,m} P^{(k)}(m,g,\delta=1)\vert_{m\equiv \tilde m} \equiv 0\;,
\label{OPT}
\ee  
determining a nontrivial optimized mass $\t m(g)$. It is consistent with renormalizability~\cite{gn2,qcd1} 
and gauge invariance~\cite{qcd1},  and (\ref{OPT}) realizes dimensional transmutation, 
unlike the original mass vanishing in the chiral limit.
In simpler ($D=1$) models this procedure is a particular case of 
``order-dependent mapping''~\cite{odm}, and was shown to converge exponentially fast for the oscillator energy 
levels~\cite{deltaconv}. \\
In most previous OPT applications, the linear $\delta$-expansion is used, $a=1$ 
in Eq.~(\ref{subst1}) mainly for simplicity. Moreover,  beyond lowest order, Eq.~(\ref{OPT}) 
generally gives more and more solutions at increasing orders, many being complex. Thus 
it may be difficult to select the right solutions, and unphysical complex ones are a burden. 
Our more recent approach\cite{rgopt1,rgopt_Lam,rgopt_alphas} crucially differs in two respects, which also drastically improve the convergence.
First, we combine OPT with renormalization group (RG)  
properties, by  requiring the ($\delta$-modified) expansion to satisfy, in addition to the OPT Eq.~(\ref{OPT}), a 
perturbative RG equation:
\be
\mu\frac{d}{d\,\mu} \left(P^{(k)}(m,g,\delta=1)\right) =0, 
\label{RG}
\ee 
where the (homogeneous) RG operator acting on a physical quantity is defined as\footnote{Our normalization is $\beta(g)\equiv dg/d\ln\mu = -2b_0 g^2 -2b_1 g^3 +\cdots$,  
$\gamma_m(g) = \gamma_0 g +\gamma_1 g^2 +\cdots$ with $b_i$, $\gamma_i$ up 
to 4-loop given in~\cite{bgam4loop}.} 
\be
\mu\frac{d}{d\,\mu} =
\mu\frac{\partial}{\partial\mu}+\beta(g)\frac{\partial}{\partial g}-\gamma_m(g)\,m
 \frac{\partial}{\partial m}\;.
 \label{RGop}
\ee 
Note, once combined with Eq.~(\ref{OPT}), the RG equation takes a reduced massless form:
\be
\left[\mu\frac{\partial}{\partial\mu}+\beta(g)\frac{\partial}{\partial g}\right]P^{(k)}(m,g,\delta=1)=0\;.
\label{RGred}
\ee
Now a crucial observation, 
overlooked in most previous OPT applications, is that after performing (\ref{subst1}),  perturbative RG invariance
is generally lost, so that Eq.~(\ref{RGred}) gives a nontrivial additional constraint, but RG invariance can only be restored for 
a unique value of the exponent $a$, fully determined by
the universal (scheme-independent) first order RG coefficients~\cite{rgopt_Lam,rgopt_alphas}:
\be
a\equiv \ga_0/(2b_0)\;.
\label{acrit}
\ee
Therefore Eqs.~(\ref{RGred}) and (\ref{OPT}) together completely fix 
{\em optimized} $m\equiv \t m$ and $g\equiv \t g$ values.
(\ref{acrit}) also guarantees that at arbitrary $\delta$ orders at least one of both the RG and OPT solutions $\t g(m)$ continuously matches  
the standard perturbative RG behaviour for $g\to 0$ (i.e. Asymptotic Freedom (AF) for QCD):
\be
\t g (\mu \gg \t m) \sim (2b_0 \ln \frac{\mu}{\t m})^{-1} +{\cal O}( (\ln \frac{\mu}{\t m})^{-2}),
\label{rgasympt}
\ee
moreover those AF-matching solutions are often unique at a given $\delta$ order for both the RG and OPT equations.
A connection of $a$ with RG anomalous dimensions/critical exponents 
had also been established previously in the  $D=3$ $\Phi^4$ model for the Bose-Einstein condensate (BEC) 
critical temperature shift by two independent OPT approaches~\cite{beccrit,bec2}.  
However, AF-compatibility and reality of solutions may easily be mutually incompatible beyond lowest order 
for optimized quantities in a given theory. 
A natural way out is to further exploit the RG freedom, considering a perturbative renormalization scheme change to attempt to 
recover both AF-compatible and real RGOPT solutions~\cite{rgopt_alphas}. 
\subsection{RGOPT for the spectral density}
To proceed with RGOPT, we first modify the perturbative series $\rho(\lambda,g)$ 
similarly to (\ref{subst1}), now clearly applied not on the original mass but on the
spectral value $\lambda\equiv |\lambda|$: 
\be
\lambda\to \lambda (1-\delta)^a\, \;\;g\to \delta\,g\, .
\label{substlam}
\ee
 and instead of (\ref{OPT}), optimizing $\rho(\lambda,g)$ with respect to $\lambda$,
\be
\frac{\partial \rho^{(k)}(\lambda,g)}{\partial \lambda}= 0\, ,
\label{OPTlam}
\ee
at successive $\delta^k$ order.
The RG equation for $\rho(g,\lambda)$ can be obtained from the defining integral 
representation of the spectral density (\ref{SD}) and the basic algebraic identity $
\partial_m \frac{m}{\lambda^2 + m^2} = 
-\partial_\lambda \frac{\lambda}{\lambda^2 + m^2}$. After some algebra 
one finds~\cite{rgopt_qq} that $\rho(\lambda)$ obeys the same RG equation as $\la \bar q q \ra$,
with  $\partial m$ replaced by $\partial \lambda$ as intuitively expected:
\be
\!\!\!\!\!\!\!\!\!\!\!
\left[\mu\frac{\partial}{\partial\mu}+\beta(g)\frac{\partial}{\partial g}-\gamma_m(g)\,\lambda
 \frac{\partial}{\partial\lambda} -\gamma_m(g) \right]\rho(\lambda,g)=0.
\label{RGlam}
\ee 
\section{Perturbative three-loop quark condensate}
We start from the standard perturbative quark condensate, calculated for non-zero quark masses.
At three-loop order in the $\ms$-scheme it reads, 
\bea
m\,\la \bar q q \ra =
 &\frac{3m^4}{2\pi^2} 
\left(\frac{1}{2}-L_m +\frac{g}{\pi^2} (L_m^2 -\frac{5}{6} L_m +\frac{5}{12}) \right.\nn \\
&\left. +(\frac{g}{16\pi^2})^2\, 
q_3(m, n_f)\right),
\label{qqQCDpert}
\eea
where $m= m(\mu)$ ($L_m\equiv\ln m/\mu$) and $g\equiv 4\pi\alpha_S(\mu)$ are the running mass and coupling in the $\ms$ scheme, 
and the three-loop coefficient $q_3(m,n_f)$ originally calculated in \cite{vac_anom3} is given in our normalization in \cite{rgopt_qq}. In
dimensional regularization (\ref{qqQCDpert}) needs 
extra subtraction after mass and coupling renormalization, namely the $m \qq$ operator has mixing with 
$m^4\times 1\!\!1$. We define~\cite{qcd1,rgopt_qq} the subtraction perturbatively as
\be
{\rm sub}(g,m) \equiv \frac{m^4}{g} \sum_{i\ge 0} s_i  g^{i},
\label{sub}
\ee
with coefficients determined order by order by requiring perturbative RG invariance.
Note that applying the RG operator (\ref{RGop}) to (\ref{sub}) defines the 
anomalous dimension of the QCD (quark) vacuum energy, explicitly known to three-loop 
order~\cite{vac_anom3,vac_anom4}.
\subsection{Perturbative spectral density}
According to Eq.~(\ref{disc}), calculating the perturbative spectral density 
formally involves calculating all logarithmic discontinuities. 
In expression (\ref{qqQCDpert}), all nonlogarithmic terms (as well as the subtraction (\ref{sub})) do not contribute, trivially giving no discontinuities,
while powers of $L_m \equiv \ln m/\mu$ follow substitution rules (\ref{disc}), e.g.
\be
L_m \to 1/2;\; 
L^2_m \to L_\lambda;\; 
L^3_m \to \frac{3}{2} L^2_\lambda -\frac{\pi^2}{8},
\label{disc3}
\ee
with $L_\lambda\equiv \ln \lambda/\mu$. 
Accordingly the QCD spectral density up to three-loop order thus reads in $\ms$-scheme
\be
\!\!\!\!\!\!\!\!\!
-\rho^{\ms}(\lambda)=  \frac{3|\lambda|^3}{2\pi^2}\,  
\left(-\frac{1}{2} +\frac{g}{\pi^2} \left(L_\lambda -\frac{5}{12}\right) +{\cal O}(g^2)\right) ,
\label{SDqq3l}
\ee
where the three-loop ${\cal O}(g^2)$
term is easily determine from 
$q_3(m,n_f)$ using (\ref{disc3}). One can then proceed with RGOPT applying Eqs.~(\ref{substlam}),(\ref{OPTlam}),(\ref{RGlam}).
Noting that 
$m \la \bar q q\ra$ is (all order) RG invariant rather than $\la \bar q q\ra(\mu)$, 
the RG-consistent value of $a$ in (\ref{substlam}) for $\la \bar q q \ra$, and the 
related spectral density from (\ref{SD}), is 
\be
a=\frac{4}{3}(\frac{\gamma_0}{2b_0}).
\ee
The OPT and RG Eqs~(\ref{OPTlam}), (\ref{RGlam}) have a first non-trivial (unique) solution at two-loop ($\delta$) order, 
given in Table~\ref{tab} for $n_f=2, 3$, using also~(\ref{BC}). We apply the RG Eq.~(\ref{RGlam}) 
consistently at two-loop order. Note that the precise number for $\la \bar q q \ra/\bar\Lam^3$ depends on the 
definition of the $\bar\Lam$ reference scale, which is a matter of convention. We adopt a four-loop order
perturbative definition~\cite{rgopt_alphas,PDG} of $\bar\Lam$, as usual in most recent analyses.
The results are conveniently given for the RG-invariant condensate: 
\be
\!\!\!\!\!\!\!\!\!\!\!
\frac{\la\bar q q\ra_{RGI}}{\la\bar q q\ra(\mu)}=(2b_0\,g)^{\frac{\gamma_0}{2b_0}}\left(1+
(\frac{\gamma_1}{2b_0} -\frac{\gamma_0\,b_1}{2b^2_0})\,g +\cdots\right),
\label{qqRGI}
\ee
\begin{table}[h!]
\begin{center}
\caption[long]{Optimized results at successive orders for $n_f=2$ and $n_f=3$, for $\t \lambda$, $\t \alpha_S$, and 
 $\la\bar q q\ra^{1/3}_{RGI}$  from~(\ref{qqRGI}). } 
\begin{tabular}{|l||c|c|c|}
\hline
$\delta^k$, RG order ($n_f=2$) &  $\ln \frac{\t \lambda}{\mu} $ & $\t \alpha_S$ & 
 $\frac{-\la\bar q q\ra^{1/3}_{RGI}}{\bar\Lam_2}$ \\
\hline
$\delta$, RG 2-loop &   $-0.45 $    & $0.480$ &  $0.821$ \\
\hline
$\delta^2$,  RG 3-loop  & $ -0.703 $ & $ 0.430 $ & $0.783$  \\
\hline
$\delta^3$, RG 4-loop  & $-0.820$ & $0.391$ &  $0.773$ \\
\hline\hline
$\delta^k$, RG order ($n_f=3$) &  $\ln \frac{\t \lambda}{\mu} $ & $\t \alpha_S$ & 
 $\frac{-\la\bar q q\ra^{1/3}_{RGI}}{\bar\Lam_3}$ \\
\hline
$\delta$, RG 2-loop &   $-0.56 $    & $0.474$ & $0.789$ \\
\hline
$\delta^2$,  RG 3-loop  & $ -0.788 $ & $ 0.444 $   & $0.766$ \\
\hline
$\delta^3$, RG 4-loop  & $-0.958$ & $0.400$ &  $0.744$ \\
\hline
\end{tabular}
\label{tab}
\end{center}
\end{table}
where higher order terms are easily derived from integrating 
$\exp{[\int\, dg \gamma_m(g)/\beta(g)]}$ at appropriate order 
using known RG function coefficients~\cite{bgam4loop}. \\
At three-loop $\alpha^2_S,\delta^2$ order, the $n_f$ dependence enters explicitly within the perturbative expression of 
the spectral density, which may thus affect the variation of the condensate value with the 
number of flavors. At higher four-loop $\alpha_S^3$ order the exact condensate expression is not known at present. But
RG recurrence properties predict~\cite{rgopt_qq} all logarithms $\ln^p m/\mu, p=1,..4$ four-loop order coefficients.
Now since only $\ln^p m$ contribute to the spectral density, the latter is fully determined
at four-loop order.  At three- and four-loop orders 
we find a unique real~\footnote{Thus for the spectral density the above mentioned more involved procedure~\cite{rgopt_alphas} 
of renormalization scheme changes to recover real optimized solutions is not required.} and unambiguously 
AF-compatible optimized solution, 
given for $n_f=2, 3$ in  Table~\ref{tab}. 
The value of $\la\bar q q\ra^{1/3}/\bar\Lam$ changes very midly as compared with the 
two-loop order result, reflecting a strong stability. 
It also appears that the {\em ratio} of 
the quark condensate to $\bar\Lam^3$ has a moderate dependence 
on $n_f$, but there is a definite trend
that $\la\bar q q\ra^{1/3}_{n_f=3}$ is smaller by about $2-3\%$ with respect to 
$\la\bar q q\ra^{1/3}_{n_f=2}$, in units of $\bar\Lambda_{n_f}$,  at the same perturbative orders. 
The stabilization/convergence is clear for the scale-invariant condensate $\la\bar q q \ra_{RGI}$ given in the last columns
in Table~\ref{tab}.
\section{Phenomenological comparison}
To better compare our results with other determinations in the literature we 
should evolve the scale-dependent condensate to the most often adopted scale $\mu=2 \rm GeV$. 
We take the scale-invariant condensate~(\ref{qqRGI}) values from the last columns of Tab.~\ref{tab} and extract from those
the condensate at any chosen (perturbative) scale $\mu'$ by using again~(\ref{qqRGI}) now taking $g \equiv 
4\pi\alpha_S(\mu')$~\footnote{For $n_f=3$ we take into account properly the charm
quark mass threshold effects~\cite{matching4l} on $\alpha_S(\mu\sim m_c)$}.
Putting all this together we obtain
\bea
&& -\la\bar q q\ra^{1/3}_{n_f=2}(2 {\rm GeV}) =  (0.833-0.845) \bar\Lam_2 \nn \\
&& -\la\bar q q\ra^{1/3}_{n_f=3}(2 {\rm GeV}) =  (0.814-0.838) \bar\Lam_3, 
\label{qqn232GeV}
\eea
where the range is from three- to four-loop results, defining our theoretical RGOPT error.
Taking for definiteness the most precise recent lattice values
of $\bar\Lam_2\simeq 331\pm 21$ (quark static potential method~\cite{LamlattVstatic14}), this gives
\be
\!\!\!\!\!\!\!
-\la\bar q q\ra^{1/3}_{n_f=2}(2 {\rm GeV},\rm lattice\, \bar\Lam_2)\simeq  278 \pm 2 \pm 18 \, {\rm MeV},
\label{qqfin2latt}
\ee
where the first error is from (\ref{qqn232GeV}) and the second from $\bar\Lam_2$ uncertainty. 
Using instead our RGOPT determination~\cite{rgopt_alphas} of $\bar\Lam_2 \simeq 360^{+42}_{-30}$ MeV 
gives somewhat higher values with larger uncertainties:
\be
\!\!\!\!\!\!\!
-\la\bar q q\ra^{1/3}_{n_f=2}(2 {\rm GeV}, \rm rgopt\, \bar\Lam_2)\simeq 301 \pm 2 ^{+35}_{-25} \, {\rm MeV}.
\label{qqfin2rg}
\ee
For $n_f=3$, using solely our RGOPT determination~\cite{rgopt_alphas} of $\bar\Lam_3= 317^{+27}_{-20}$ MeV, 
gives:
\be
\!\!\!\!\!\!\!
-\la\bar q q\ra^{1/3}_{n_f=3}(2 {\rm GeV}, \rm rgopt\, \bar\Lam_3)
\simeq  262 \pm 4 ^{+22}_{-17} \, {\rm MeV}, 
\label{qqfin3rg}
\ee
where again the errors are respectively from (\ref{qqn232GeV}) and $\bar\Lam_3$ uncertainty. \\
Rather than fixing the scale from $\bar\Lambda$, one may alternatively 
give results for the ratio of the scale-invariant condensate with another physical scale. 
Using solely RGOPT results~\cite{rgopt_alphas} for $F/\bar\Lam_2$ 
and $F_0/\bar\Lam_3$ (where $F$ ($F_0$)
are the pion decay constant for $n_f=2$, $n_f=3$ respectively in the chiral limit), we obtain
\be
\frac{\la\bar q q\ra^{1/3}_{RGI,n_f=3}}{\la\bar q q\ra^{1/3}_{RGI,n_f=2}} \simeq 
\left(0.94 \pm 0.01 \pm 0.12\right)\:\frac{F_0}{F},
\label{qq3to2}
\ee
where errors are combined linearly. In (\ref{qq3to2})
the first error is the RGOPT error for the condensate, and the second
larger one is propagated from the $F/\bar\Lam_2$ and $F_0/\bar\Lam_3$ RGOPT errors. \\
One can compare~(\ref{qqfin2latt}),~(\ref{qqfin2rg}) 
with the latest most precise lattice determination, from the spectral 
density~\cite{SDlatt_recent} for $n_f=2$:
$\la\bar q q\ra^{1/3}_{n_f=2}(\mu=2 {\rm GeV}) =-(261 \pm 6 \pm 8) $, where the first error is statistical 
and the second is systematic. Our result (\ref{qqfin2latt}) is thus compatible within uncertainties.
For $n_f=3$ the most precise lattice  determination we are aware of is $\la\bar q q\ra^{1/3}_{n_f=3}(2\rm GeV) =
-(245\pm 16)$ MeV~\cite{Lattqqn3}.
Our results compare also well with the latest ones from spectral sum 
rules~\cite{qqSRlast}: $\la\bar u u\ra^{1/3} \sim -(276\pm 7)$ MeV. 
\section{Summary and Conclusion}
Our recent RGOPT determination~\cite{rgopt_qq} of the quark condensate via the 
spectral density of the Dirac operator gives successive sequences of nontrivial optimized results 
in the chiral limit. At two-, three- and four-loop levels it exhibits
a remarkable stability. The intrinsic theoretical error, 
taken as the difference between three- and four-loop results, is of order $2\%$. 
The final condensate value uncertainty is more affected by the present uncertainties 
on the basic QCD scale $\bar\Lam$, with a larger uncertainty for $n_f=2$ flavors. 
The values obtained are rather compatible, within uncertainties, with the most recent lattice 
and sum rules determinations for $n_f=2$, and indicate a moderate flavor dependence of 
the $\la\bar q q\ra^{1/3}_{n_f}/\bar\Lam_{n_f}$ ratio, in some contrast with the results in \cite{qqflav}.
Since our results are by construction valid in the strict chiral limit, 
they indicate that the possibly larger difference obtained by some other determinations is more likely due to the
explicit breaking from the large strange quark mass, rather than a large intrinsic $n_f$ dependence  
of the condensate in the exact chiral limit. 


\begin{thebibliography}{999}
%
\bibitem{GMOR} M.~Gell-Mann, R.~J.~Oakes and B.~Renner,
   Phys.\ Rev.\  {\bf 175}, 2195 (1968).
%
\bibitem{LattFLAG} G.~Colangelo {\it et al.}, Eur. Phys. J. C {\bf 71}, 1695 (2011);
S. Aoki {\it et al.}, arXiv:1310.8555 [hep-lat].
%
\bibitem{SVZSSR}
M.~A.~Shifman, A.~I.~Vainshtein, and V.~I.~Zakharov,  Nucl.\ Phys.\ {\bf B147}, 385 (1979).
%
\bibitem{qqSR}  See {\it e.g.} H.~G.~Dosch and S.~Narison,
  Phys.\ Lett.\ B {\bf 417}, 173 (1998); 
   M. Jamin, 
   Phys.\ Lett.\ B {\bf 538}, 71 (2002).
%
\bibitem{qqSRlast} 
S.~Narison,
  Phys.\ Lett.\ B {\bf 738}, 346 (2014).
%
\bibitem{NJL} Y.~Nambu and G.~Jona-Lasinio,
  Phys.\ Rev.\  {\bf 122}, 345 (1961).
%
\bibitem{NJLrev} S.~P.~Klevansky,
  Rev.\ Mod.\ Phys.\  {\bf 64}, 649 (1992);
  T.~Hatsuda and T.~Kunihiro,
  Phys.\ Rept.\  {\bf 247}, 221 (1994).
%
  \bibitem{D-S} P.~Maris, C.~D.~Roberts, and P.~C.~Tandy, Phys.\ Lett.\ B {\bf 420}, 267 (1998);
 P.~Maris and C.~D.~Roberts, Phys.\ Rev.\ C {\bf 56}, 3369 (1997).
%
\bibitem{qqbar_link} K.~Langfeld, H.~Markum, R.~Pullirsch, C.~D.~Roberts, and S.~M.~Schmidt,
  Phys.\ Rev.\ C {\bf 67}, 065206 (2003).
%
\bibitem{qqlattSD}
  L.~Del Debbio, L.~Giusti, M.~L\"uscher, R.~Petronzio, and N.~Tantalo,
  J. High Energy Phys. {\bf 0602}, 011 (2006);
 L.~Giusti, and M.~L\"uscher, J.\ High Energy Phys. 03 ({\bf 2009}) 013;
 H.~Fukaya, S.~Aoki, T.W.~Chiu, S.~Hashimoto,
T.~Kaneko, J.~Noaki, T.~Onogi, and N.~Yamada, Phys.\ Rev.\ Lett.\ {\bf 104}, 122002 (2010) [Erratum-{\it ibid.}\  {\bf 105}, 159901 (2010)].
%
\bibitem{SDlatt_recent}
G.~P.~Engel, L.~Giusti, S.~Lottini and R.~Sommer,
  Phys.\ Rev.\ D {\bf 91}, 054505 (2015)
%
\bibitem{BanksCasher}   T.~Banks and A.~Casher, Nucl.\ Phys.\ {\bf B169}, 103 (1980);
%
\bibitem{SDgen} H.~Leutwyler and A.~V.~Smilga,
  Phys.\ Rev.\ D {\bf 46} (1992) 5607.
\bibitem{SDchpt2}  
  A.~V.~Smilga and J.~Stern,
  Phys.\ Lett.\ B {\bf 318}, 531 (1993);
  K.~Zyablyuk,  J. High Energy Phys. {\bf 0006}, 025 (2000).
%
\bibitem{chpt}
J.~Gasser and H.~Leutwyler,
  Annals\ Phys.\  {\bf 158}, 142 (1984);
Nucl.\ Phys.\ {\bf B250}, 465 (1985).
%
\bibitem{qqflav} See e.g.
V.~Bernard, S.~Descotes-Genon, and G.~Toucas,
  J. High Energy Phys. {\bf 1206}, 051 (2012); and earlier references therein.
%
\bibitem{rgopt1} J.-L.~Kneur and A.~Neveu, Phys.\ Rev.\ D {\bf 81}, 125012 (2010). 
%
\bibitem{rgopt_Lam}  J.-L.~Kneur and A.~Neveu,
  Phys.\ Rev.\ D {\bf 85}, 014005 (2012).
%
\bibitem{rgopt_alphas}  J.-L.~Kneur and A.~Neveu,
 Phys.\ Rev.\ D {\bf 88}, 074025 (2013).
%
\bibitem{rgopt_qq} J.~L.~Kneur and A.~Neveu,
  Phys.\ Rev.\ D {\bf 92},  074027 (2015).
%
%
\bibitem{delta} There are numerous references on the delta-expansion, see
{\it e.g.} ref. [21] in \cite{rgopt_alphas}.  
%
\bibitem{gn2} C.~Arvanitis, F.~Geniet, M.~Iacomi, J.-L.~Kneur, and A.~Neveu,
Int.\ J.\ Mod.\ Phys.\ A {\bf 12}, 3307 (1997).
%
\bibitem{qcd1} C.~Arvanitis, F.~Geniet, J.~L.~Kneur, and A.~Neveu,
  Phys.\ Lett.\ B {\bf 390}, 385 (1997);
  J.-L. Kneur, Phys.\ Rev.\ D {\bf 57}, 2785 (1998).
%
\bibitem{odm} R. Seznec and J. Zinn-Justin, J. Math. 
 Phys. {\bf 20}, 1398 (1979); J.C. Le Guillou and J. Zinn-Justin,
Ann. Phys. {\bf 147}, 57 (1983); J. Zinn-Justin, arXiv:1001.0675.
%
\bibitem{deltaconv} R.~Guida, K.~Konishi, and H.~Suzuki, Ann. Phys. (N.Y.) {\bf 241}, 152
(1995); {\bf 249}, 109 (1996). 
%
\bibitem{bgam4loop} 
  J.~A.~M.~Vermaseren, S.~A.~Larin, and T.~van Ritbergen,
  Phys.\ Lett.\ B {\bf 405}, 327 (1997).
K.~G.~Chetyrkin, Nucl.\ Phys.\ {\bf B710}, 499 (2005);
M.~Czakon Nucl.\ Phys.\ {\bf B710}, 485 (2005).
%
\bibitem{beccrit} H.~Kleinert, Mod.\ Phys.\ Lett.\ B {\bf 17}, 1011 (2003);
 B.~Kastening, Phys.\ Rev.\ A {\bf 68}, 061601 (2003); Phys.Rev. A {\bf 69}, 043613 (2004);
B.~Hamprecht and H.~Kleinert,
  Phys.\ Rev.\ D {\bf 68}, 065001 (2003).
%
\bibitem{bec2} J.-L.~Kneur, A.~Neveu, and M.~B.~Pinto,
Phys.\ Rev.\ A {\bf 69}, 053624 (2004).
%
\bibitem{vac_anom3}
K.~G.~Chetyrkin and J.~H.~K\"uhn, Nucl.\ Phys.\ {\bf B432}, 337 (1994);
K.~G.~Chetyrkin and A.~Maier,  J.\ High\ Energy\ Phys.\ 01 ({\bf 2010}) 092.
%
\bibitem{vac_anom4} K.~G.~Chetyrkin and A.~Maier, private communication.
%
\bibitem{PDG} K.~A.~Olive {\it et al.}  [Particle Data Group Collaboration],
  Chin.\ Phys.\ C {\bf 38}, 090001 (2014).
%
\bibitem{matching4l} See e.g. 
K.~G.~Chetyrkin, J.H K\"uhn and C.~Sturm,  Nucl.\ Phys.\ {\bf B744}, 121 (2006).
%
\bibitem{LamlattVstatic14} F.~Karbstein, A.~Peters and M.~Wagner,
  JHEP {\bf 1409}, 114(2014).
%
\bibitem{Lattqqn3} A. Bazavov {\it et al.}, 
[arXiv:0910.2966].
%
\end{thebibliography}
\end{document}